\documentclass[a4paper]{jpconf}
\usepackage{graphicx}
\usepackage{amsmath,amssymb}
\usepackage{geometry}
\usepackage{color}
\DeclareGraphicsExtensions{.ps,.eps,.EPSF} 
\begin{document}
\newcommand{\spd}{$sp$--$d$ }
\newcommand{\ef}{E_{\rm F}}
\newcommand{\du}{{\rm d}}
\newcommand{\Ang}{{\rm \AA}}
\newcommand{\be}{\begin{equation}}
\newcommand{\ee}{\end{equation}}
\newcommand{\ben}{\begin{eqnarray}}
\newcommand{\een}{\end{eqnarray}}
\newcommand{\beq}{\begin{equation}}
\newcommand{\eeq}{\end{equation}}
\newcommand{\B}{\mathrm{B}}
\newcommand{\NB}{\mathrm{NB}}
\newcommand{\RRe}{\mathrm{Re}}
\newcommand{\IIm}{\mathrm{Im}}
\title{Sub-diffusive electronic states in octagonal tiling}
\author{G. Trambly de Laissardi\`ere$^{\dagger}$, C. Oguey$^{\dagger}$ and D. Mayou$^{\ast \ddagger}$}
\address{
$^{\dagger}$ Laboratoire de Physique th\'eorique et Mod\'elisation, CNRS and 
Universit\'e de Cergy-Pontoise, 
95302 Cergy-Pontoise, France \\
$^{\ast}$ Univ. Grenoble Alpes, Inst NEEL, F-38042 Grenoble, France \\
$^{\ddagger}$ CNRS, Inst NEEL, F-38042 Grenoble, France
}
\ead{guy.trambly@u-cergy.fr}
\begin{abstract}
We study the quantum diffusion of charge carriers in octagonal tilings. 
Our numerical results show a power law decay of the wave-packet spreading, 
$L(t) \propto t^{\beta}$, characteristic of critical states in quasicrystals at large time $t$.
For many energies states are sub-diffusive, i.e. $\beta < 0.5$, and thus conductivity increases when the amount of defects (static defects and/or temperature) increases. 
\end{abstract}

\vskip -.2cm
Experimental investigations have indicated that the conduction
properties of many stable quasicrystals (AlCuFe, AlPdMn, AlPdRe, ...) are unusual and differ strongly 
from those of simple inter-metallic alloys
[1-3].
In particular their conductivity increases with static defects density and when temperature increases. 
It appears also that the medium range order and the chemical order --over one or a few nanometers--
have a decisive influence
[4-9].
There is now strong evidence that these non standard properties
result from a new type of break-down of the semi-classical Bloch-Boltzmann theory of conduction
[10-14].
On the other hand, the specific role of long range quasiperiodic order in electronic properties is
still an open question in spite of a large number of studies (Refs. 
[15-37]
and Refs. therein).
Many studies support the existence of critical states, which are neither extended nor localised, but are characterised by a power law decay of the wave-function envelope at large distances. 
In the presence of critical states, the diffusion of charge carrier at sufficiently large time $t$ follows a power law and then the spatial extension $L$ of wave-packets should be, 
[34-37]
\vskip -.2cm 
\begin{equation}
L(E,t)  \propto t^{\beta(E)} ~{\rm at~large~} t,
\label{eq_Lt}
\end{equation}
where $\beta$, $0 \le \beta \le 1$, is an exponent depending on energy $E$ and on the Hamiltonian model. 
Note that in usual metallic crystals without static defects, $\beta = 1$ and the propagation is ballistic.
In strongly disordered systems, $\beta =0.5$ for a large time range and propagation is diffusive. 
For a localised state one has $\beta = 0$. 
When disorder is introduced in the perfect approximant or perfect quasicrystal in the form of static defects (elastic scatterers) and/or inelastic scattering (temperature, magnetic field...), the defects induce scattering and we expect that there is an associated time $\tau$ above which the propagation 
of the wave-packet is diffusive. 
The diffusivity $D$ of charge carrier at energy $E$ can be estimated by,
$D(E,\tau) \simeq {L(E,t=\tau)^2}/{\tau} \propto \tau^{2 \beta(E) - 1}$,
and the conductivity $\sigma$ at zero frequency is given by the Einstein formula: 
\vskip -.2cm 
\begin{equation}
\sigma(E_F,\tau) = {\rm e}^2 n(E_F) D(E_F,\tau) \propto \tau^{2\beta(E_F)-1},
\end{equation} 
where $n$ is the density of states and $E_F$ the Fermi energy. 
The case $0.5 < \beta < 1$, called super-diffusive regime, leads to transport properties similar to metal,
since the conductivity decreases when disorder increases --i.e. when $\tau$ decreases--.
Conversely, for $0 < \beta < 0.5$, the regime is sub-diffusive and the conductivity increases when disorder increases like in real quasicrystals. 
Many authors consider 
[18-20,29,34-37]
that critical states could lead to $\beta < 0.5$ but it has not yet been shown in 2D or 3D quasiperiodic structures 
(except for some very specific energies). 

{\it Model Hamiltonian.}--
The octagonal, or Ammann-Beenker, tiling \cite{Ammann92,Socolar89} 
is a quasiperiodic tiling analogous to the notorious Penrose tiling.
This tiling 
has been often used to understand the influence of quasiperiodicity on electronic transport
[18-25].
A sequence of periodic approximants $X_0, X_1,\ldots,X_k,\ldots$ can be generated
\cite{DuMoOg89}.
In approximants of order 
$k\geq 1$, the 6 local configurations around vertexes are the same as in the octagonal quasiperiodic tiling. 
They have, respectively, coordination number $\eta = 3, 4, 5, 6, 7$ and $8$. 
We consider the simple Hamiltonian,
\vskip -.1cm 
\begin{equation}
\hat{H} = \sum_{i} \epsilon_i c_i^*c_i + \sum_{\langle i;j \rangle} \gamma c_i^*c_j + h.c. ,
\label{Eq_hamiltonian}
\end{equation}
where $i$ indexes s orbitals located on vertexes, and $\gamma$ 
is the strength of the hopping between orbitals.
$\langle i,j \rangle$ are the nearest-neighbours at tile edge distance $a$. 
To simulate schematically a possible effect of the presence of different chemical elements,
the one-site energy $\epsilon_i$ is proportional to the coordinance $\eta_i$ of the site $i$: 
$\epsilon_i = \eta_i \gamma $.
To obtain realistic time values, we use $\gamma = 1$\,eV which is the order of magnitude of 
the hopping parameter in real inter-metallic compounds.
The total density of states (total DOS) of $X_7$ approximant is shown figure 1(a).

{\it Quantum diffusion.}--
In the framework of Kubo-Greenwood approach 
for calculation of the conductivity,
we use the polynomial expansion method developed by Mayou, Khanna, Roche and 
Triozon \cite{Mayou88,Mayou95,Triozon02,Roche97,Roche99} 
to compute the mean square spreading of the wave-packet at time $t$ and energy $E$: 
$L^2(E,t) = \langle (\hat{X}(t)-\hat{X}(0) )^{2} \rangle_{E}$, 
where $\hat{X}$ is the position operator in the x-direction.
\begin{figure}
\includegraphics[width=7.1cm]{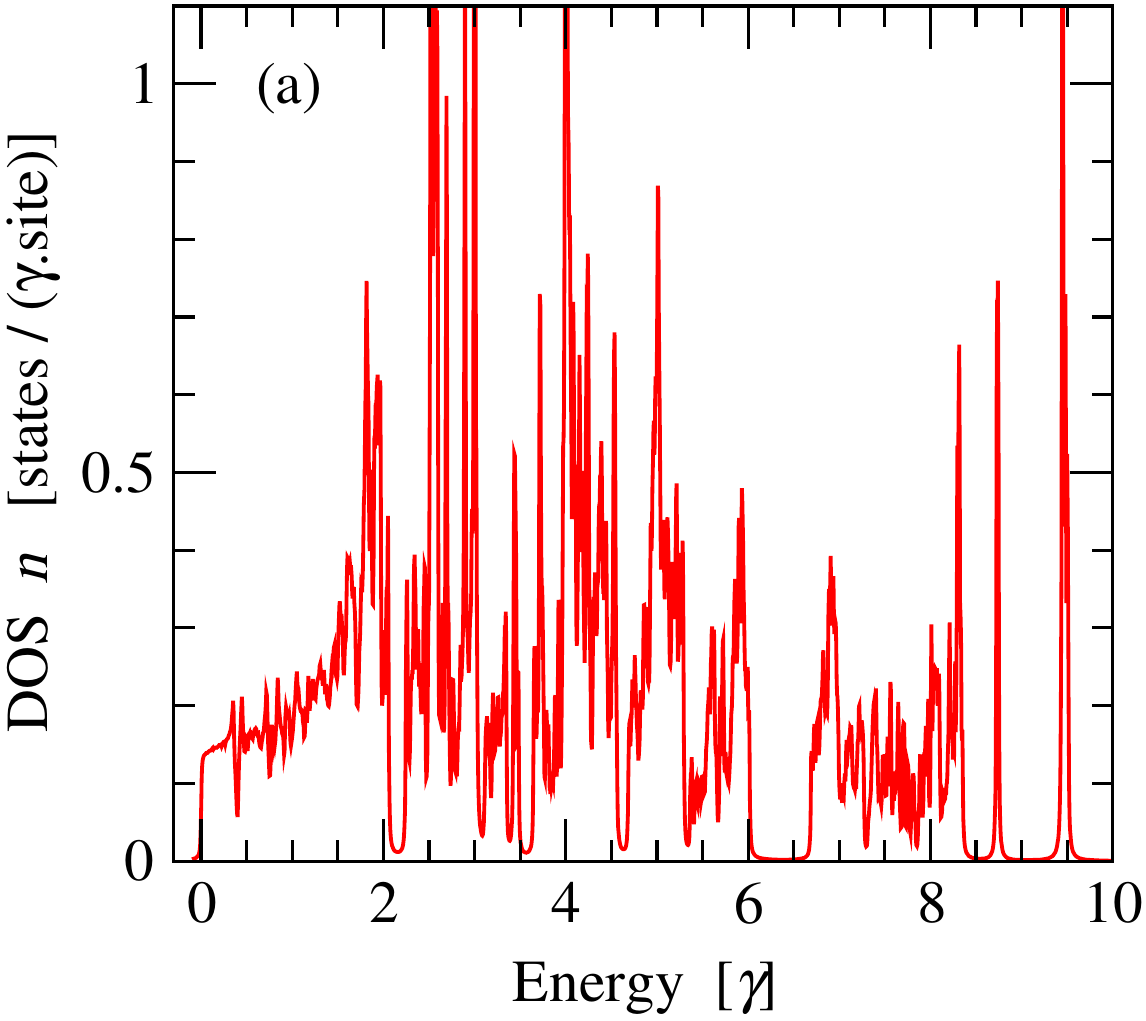}
~\includegraphics[width=7.4cm]{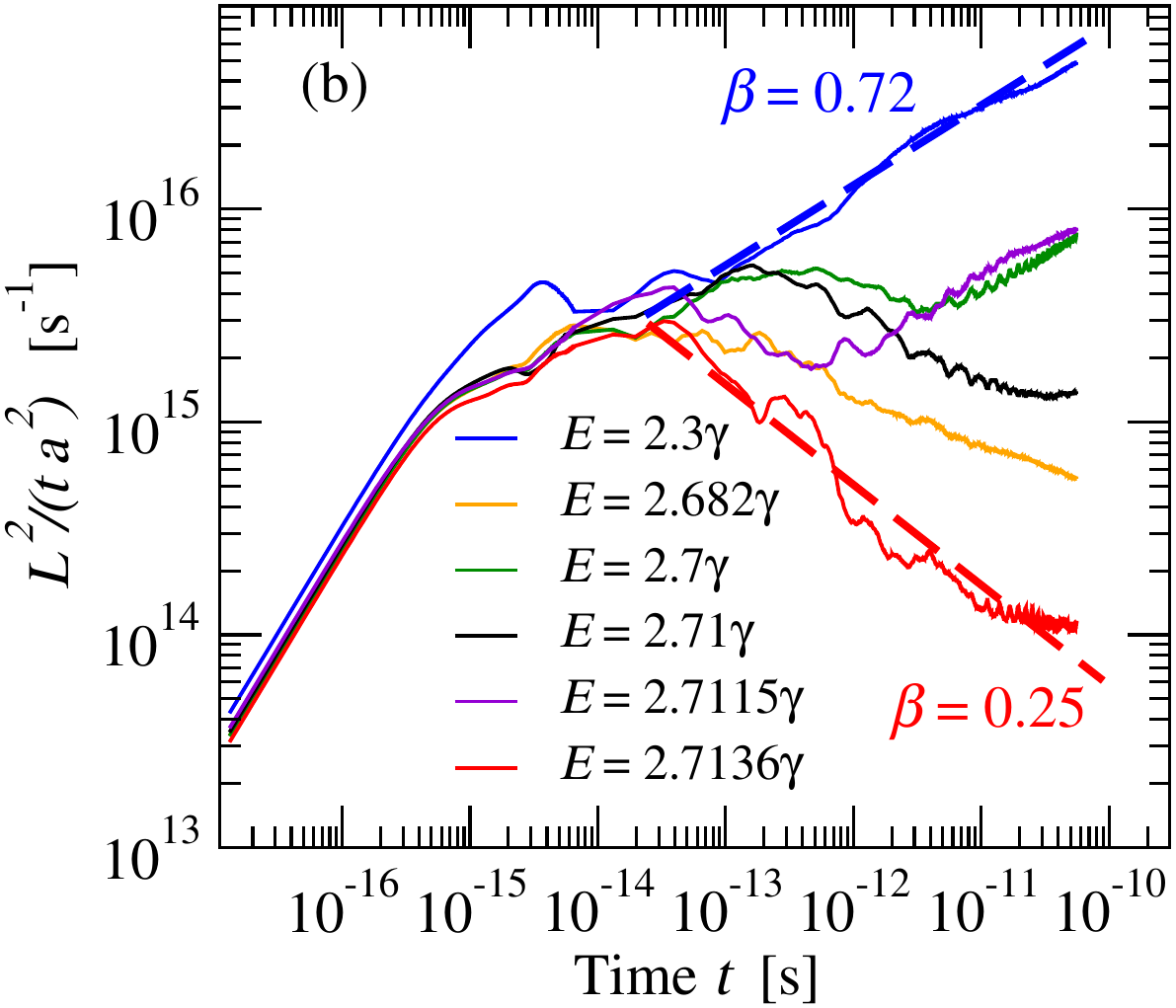}
\caption{ 
Density of states and quantum transport in perfect $X_7$ octagonal approximants  
(number of sites per cell $N = 275807$, unit cell $478a$$\times$$478a$, with $a$ the tile edge length):
(a) Total DOS $n$ (the recursion method induces a convolution of $n(E)$ by a Lorentzian of width 5\,meV), 
(b) diffusive coefficient $\mathcal{D}(t) = L(t)^2 / t$ for various $E$.
}
\label{Fig_Dt}
\end{figure}
The diffusion coefficient $\mathcal{D}(E,t) = L(E,t)^2 / t$ 
is shown in figure \ref{Fig_Dt}(b) for $X_7$ approximant
at some energies. 
The ballistic regime due to the periodicity of the approximant is reached at very large $t$, 
when $L(t) > L_k$ where $L_k$ is the approximant cell size;
then $L(t) = V_B t$, where $V_B$ is the Boltzmann velocity, 
{$i.e.$ the intra-band velocity, $V_B = \langle \partial E_n(k)/ \partial k_x \rangle_E / \hbar$, where $E_n(k)$ is the  
band dispersion relation \cite{Trambly14CRAS}.} 
For $X_7$ in time range
shown figure 1,
this Boltzmann term is negligible and, for all purposes of this discussion, 
the $X_7$ approximant is equivalent to the quasiperiodic system.
Depending on the $t$ values, three different regimes are observed at each energy:
\begin{itemize}
\item
At very small time, typically when $L(t) < a$, the mean spreading grows linearly with $t$,  
$L(t) = V_0 t$ (ballistic behaviour), where $V_0 > V_B$ \cite{PRL06,Trambly14CRAS}.
\item 
For times, corresponding to $L(t) \gtrsim {\rm a~few~}a $, the propagation seems to become diffusive as the diffusion coefficient is almost constant, $\mathcal{D}(t) \simeq \mathcal{D}_{dif}$. 
Therefore $L_1$, defined by
$L_1 = {\mathcal{D}_{dif}}/{V_0} \simeq {\rm a~few~}a $,
is a kind of effective elastic scattering length but it is not due to static scattering events because we consider perfect tilings. 
The corresponding effective elastic scattering time is $t_1 = L_1 / V_0$. 
Roughly speaking, it seems that when $L(t) \gtrsim L_1$,
$i.e.$ $t \gtrsim t_1$, the wave-packet feels a random tiling.
\item 
An other distance $L_2$ (respectively an other time $t_2$, $L(t_2) = L_2$) appears.
For $L(t) > L_2 \simeq {\rm a~few~}10a$, a new regime appears and $\mathcal{D}(t)$ follows a power law.
It is thus characteristic of the medium and long range quasiperiodic order. 
Figure 1(b) shows that the $\beta$ value can switch from a sub-diffusive regime ($\beta < 0.5$) to a super-diffusive regime ($\beta > 0.5$) over a small variation of energy.
The $t_2$ values, $t_2 \simeq 10^{-13}$--${10^{-14}}$\,s, have the order of magnitude of the scattering time above which measurements show unusual transport properties in quasicrystals \cite{Berger94}.
\end{itemize}
Both distances $L_1 \simeq {\rm a~few~} a$, $L_2 \simeq {\rm a~few~} 10a$, and the exponent $\beta$ at time $t> t_2$, depend a lot on the energy value $E$. 
$L_1 < L_2$, but at some energy it even seems that $L_1 \simeq L_2$. 
Further analysis are necessary to understand the energy dependence. 

To summarise, we have presented quantum diffusion in a large approximant 
of the octagonal tiling.
The charge carrier propagation is determined by the wave-packet spreading in the quasiperiodic lattice. 
From numerical calculation, two length scales seem to characterise this quasiperiodic spreading. 
$L_1$, typically $L_1 = {\rm a~few~} a$, above which the propagation is almost diffusive in spite of the absence of static defects. 
$L_2$, typically $L_2 = {\rm a~few~} 10a$, above which specific quasiperiodic symmetries lead to a 
power law dependence of the root mean square spreading, $L(t) \propto t^{\beta}$. For some energies
states are super-diffusive or diffusive, $i.e.$ $\beta \ge 0.5$, 
whereas for other energies, a sub-diffusive regime, $i.e.$ $\beta < 0.5$, sets in as expected for critical states characteristic of quasiperiodicity. 
This sub-diffusive regime is the generalisation to quasicrystal of the non-Boltzmann propagation found in
realistic approximants of i-AlMnSi and i-AlCuFe \cite{PRL06,Trambly08}, in the complex inter-metallic alloys $\lambda$-AlMn \cite{Trambly14CRAS}, and in small approximants of octagonal and Penrose tilings \cite{Trambly11,Trambly14}.

The computations were performed at the
Centre de Calcul (CDC), 
Universit\'e de Cergy-Pontoise.
We thank Y. Costes for computing assistance.

\end{document}